\documentclass[twocolumn]{aastex631}
\usepackage{bm}
\usepackage{amsmath}
\graphicspath{{./}{figures/}}

\begin{document}

\title{Polar circumtriple planets and disks can only form close to a triple star}

\author[0000-0003-2270-1310]{Stephen Lepp}

\author[0000-0003-2401-7168]{Rebecca G. Martin}

\affiliation{Nevada Center for Astrophysics, University of Nevada, Las Vegas, 4505 S. Maryland Pkwy., Las Vegas, NV 89154, USA}
\affiliation{Department of Physics and Astronomy,University of Nevada, Las Vegas, 4505 S. Maryland Pkwy., Las Vegas, NV 89154, USA}

\author[0000-0002-4636-7348]{Stephen H. Lubow}

\affiliation{Space Telescope Science Institute, 3700 San Martin Drive, Baltimore, MD 21218, USA}

\begin{abstract}
Observations of protoplanetary disks around binary and triple star systems suggest that misalignments between the orbital plane of the stars and the disks are common. Motivated by recent observations of polar circumbinary disks, we explore the possibility for polar circumtriple disks and therefore polar circumtriple planets that could form in such a disk. With $n$-body simulations and analytic methods we find that the inclusion of the third star, and the associated apsidal precession, significantly reduces the radial range of polar orbits so that circumtriple polar disks and planets can only be found close to the stellar system. 
Outside of a critical radius, that is typically in the range of $3-10$ times the outer binary separation depending upon the binary parameters, the orbits behave the same as they do around a circular orbit binary.  
{ For some observed systems that have shorter period inner binaries, the critical radius is considerably larger.}
If polar circumtriple planets can form, we suggest that it is likely that they form in a disk that was subject to breaking. 
\end{abstract}


\section{Introduction}

Multiple stellar systems are common in star forming regions \citep{Duchene2013}. Disks around triple star systems are also expected to be common \citep{Tobin2016,Bate2018} and there are several well known examples including GG Tauri A \citep{DiFolco2014,Keppler2020,Phuong2020} and GW Ori \citep{Bi2020,Kraus2020,Smallwood2021}. A common feature of these disks is that they are tilted with respect to the orbital plane of the stars. Disk misalignment may initially occur, for example, because of turbulence in the molecular gas cloud \citep{Offner2010, Tokuda2014, Bate2012} or later accretion of material by the young binary \citep{Bate2010, Bate2018}. Misalignment may be increased later by stellar flybys \citep{Nealon2020} or bound stellar companions \citep[e.g.][]{Martin2017,Martin2022}.

Around an eccentric binary star system, test particle orbits have two stable stationary states: coplanar alignment to the binary orbit, and polar alignment in which the angular momentum of the particle orbit is aligned to the binary eccentricity vector and $90^\circ$ to the binary orbital plane \citep{Verrier2009,Farago2010,Doolin2011,Chen2019}. A particle that is misaligned from one of these two stationary states undergoes nodal precession. Low initial inclination orbits precess about the binary angular momentum vector while high initial inclination orbits precess about the binary eccentricity vector.  Since the test particle does not affect the dynamics of the binary, the qualitative behaviour does not depend on orbital radius of the particle around the binary unless general relativity or tides become important \citep{Lepp2022}.

A circumbinary disk with a low-mass can undergo similar dynamical behaviour to a test particle \citep[e.g][]{Aly2015,Martin2018}. If the disk is in good radial communication, it can undergo solid body precession at a { angular momentum} weighted average { rate} \citep{Papaloizou1995,Larwoodetal1996}.  For protoplanetary disks, the radial communication is wave-like \citep{Papaloizou1983,Lubow2001}. Dissipation in the disk leads to alignment either towards coplanar \citep{Nixon2013,Facchini2013} or polar depending on the initial tilt \citep{Martin2017,Lubow2018,Zanazzi2018,Cuello2019}. Several polar circumbinary disks around eccentric binaries have been observed \citep{Kennedy2012,Kennedy2019,Kenworthy2022}, although none have yet been observed around a triple star. While polar circumbinary planets have not yet been observed, their formation may be as efficient as in a coplanar configuration \citep{Childs2021,Childs2021b}.

While the evolution of circumbinary particles and disks is now fairly well understood, the inclusion of an inner hierarchical triple star system has not been explored in detail. In this work, for the first time, we examine the effect of an inner triple star system on the existence of polar orbits.  In Section~\ref{sec:intro} we use $n$-body simulations and in Section~\ref{analytic} we compare to analytic models. The inner and outer binaries that compose the triple star undergo apsidal precession. We show that this can remove the possibility of polar orbits outside of a critical radius from the triple star. This is similar to the effects of general relativity that also causes apsidal precession of the binary \citep{Lepp2022} but with much higher precession rates. In Section~\ref{conc} we draw our conclusions and discuss implications both for observations of circumtriple disks and for the properties of planets that may form in such disks.

\section{Circumtriple particle orbits} 
\label{sec:intro}

%fig1as fig1s is from fig2bs on computer. they are in /Users/lepp/Desktop/3star/circumtriple
\begin{figure*}
\includegraphics[width=\columnwidth]{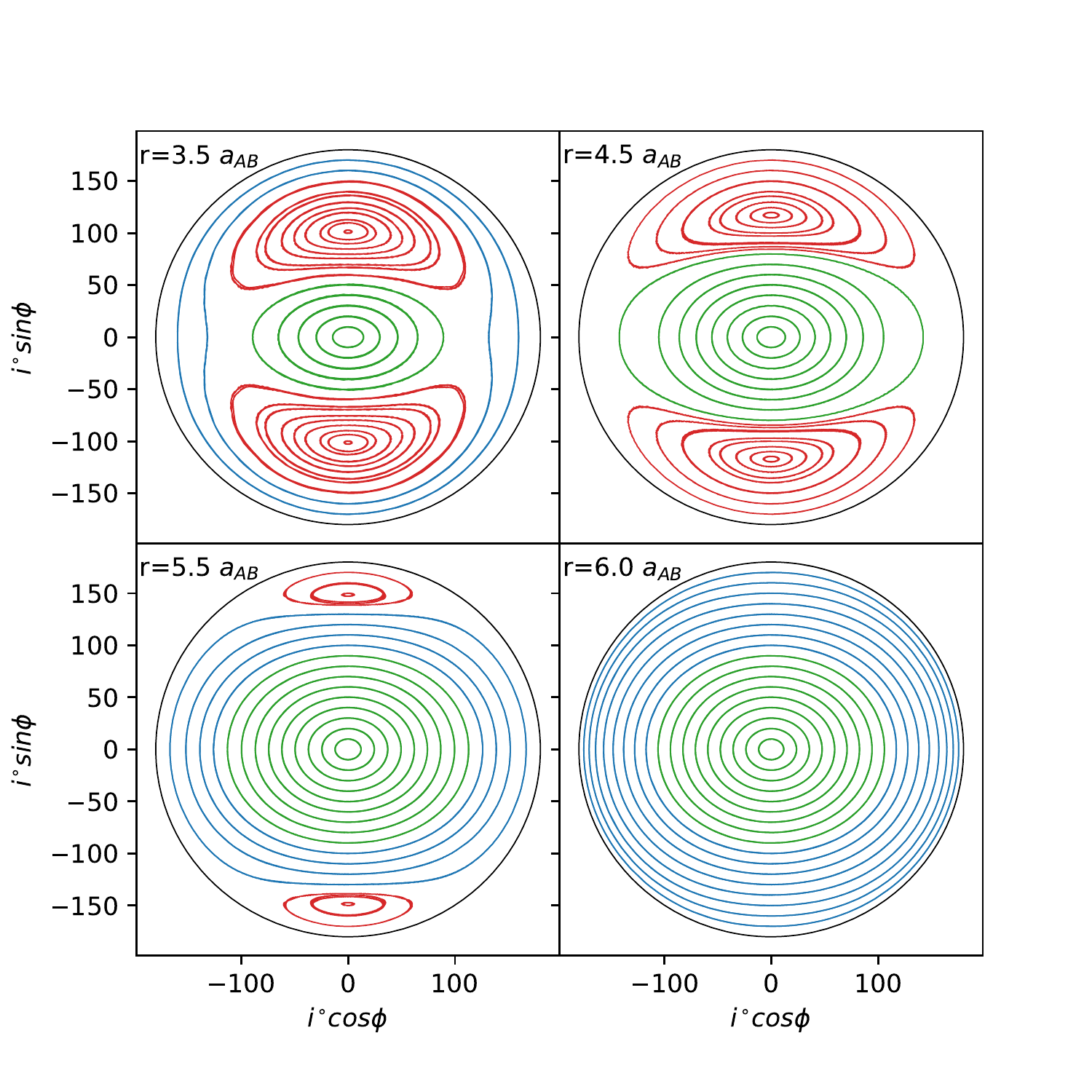}
\includegraphics[width=\columnwidth]{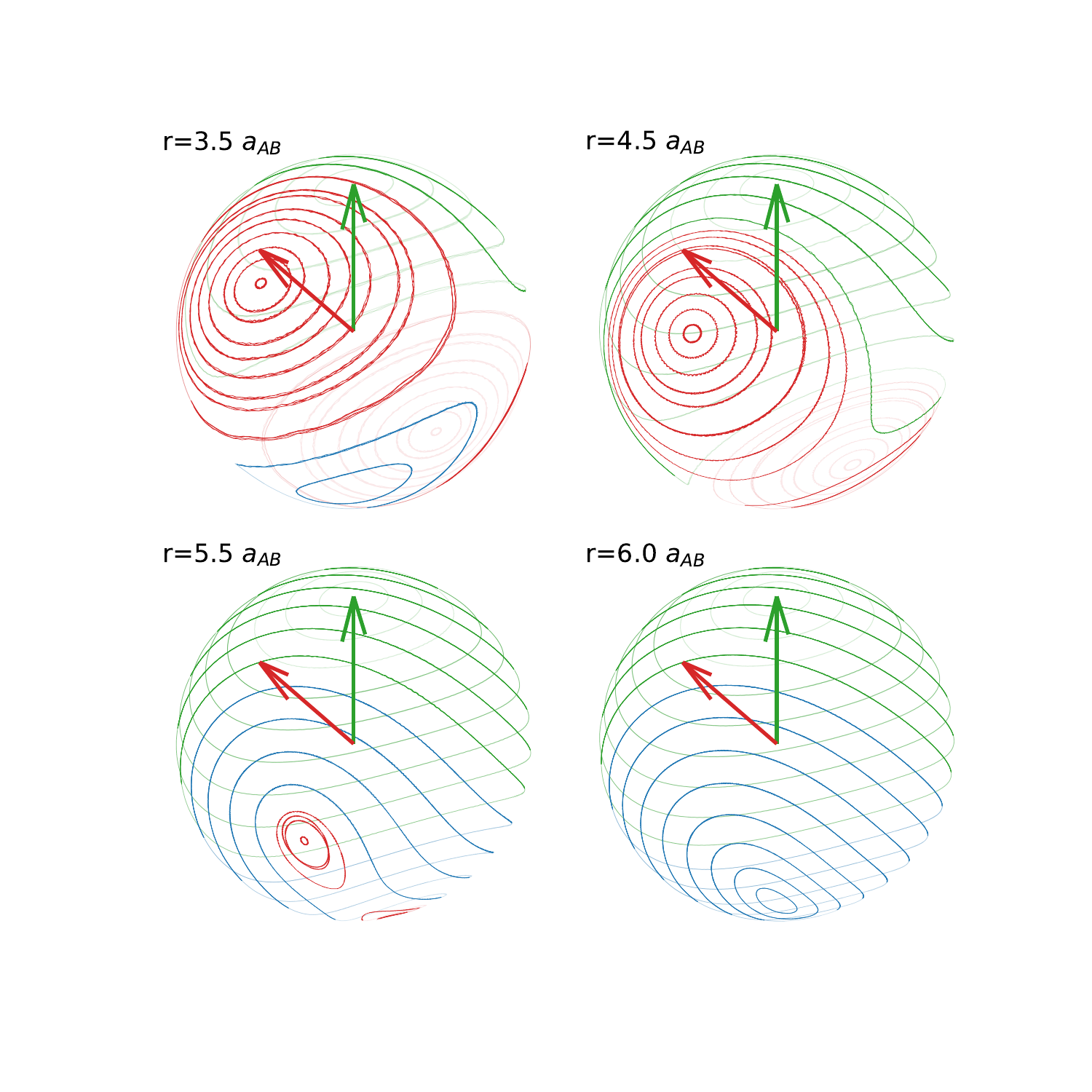}
\caption{Test particle orbits around our standard triple star  at orbital radii of $r=3.5$, 4.5, 5.5 and $6\,\rm a_{AB}$. Left: The $(i\cos\phi,i\sin\phi)$ phase plane.  Right: Precession paths for the angular momentum vector of the test particles plotted on surface of sphere. The angular momentum and eccentricity unit vectors of binary 
shown in green and red, respectively.  The circulating orbits are shown in  green, librating orbits are red and retrograde circulating orbits are blue.
}
\label{fig:phase}
\end{figure*}

In this Section we first consider the dynamics of a particle orbiting a triple star with our standard parameters and then we consider the effect of varying different triple star parameters. We use the REBOUND N-body code \citep{rebound}. The simulations were integrated using a combination of  IAS15, a 15th order Gauss-Radau integrator \citep{reboundias15} and the WHFast, a symplectic Wisdom-Holman integrator \citep{reboundwhfast,wh}.

\subsection{Triple star parameters}
\label{sec:params}

{Triple star systems are found to occur with a large range of properties.  Figure~3 of \cite{Tokovinin2021} plots the outer binary period as a function of inner binary period for a sample of 1820 systems that lie within a distance of 200pc. 
The sample is subject to strong selection effects that favor the detection of close spectroscopic binaries and resolved wide binaries. Nonetheless, the plot suggests that for longer period inner binaries ($ > 1y$),
the outer to inner semi-major axis ratios typically range from 3 to 50. For shorter period inner binaries  ($ < 1y$), the ratio range is typically from 20 to 100, and ratios of greater than 1000 also occur.}

Triple star systems may be unstable for a wide range of parameter space \citep{Mardling2001,Valtonen2006,vynatheya2022}. We consider hierarchical triple systems composed of an inner binary with an outer binary companion. The inclination of the inner binary to the inclination of the binary companion must be small enough to avoid von Zeipel-Kozai-Lidov (ZKL) oscillations \citep{Zeipel1910,Kozai1962,Lidov1962,Naoz2016,Hamers2021}. 
{Figure~3 of \cite{Tokovinin2021}  shows evidence of the stability limit at a period ratio of about 4.7 predicted by \cite{Mardling2001}. }
{For the range of parameters studied, the eccentricity of all the particle orbits
are relatively constant.}

The orbits are scale free in mass and length and we adopt as
our mass unit, the total mass of the triple star system, $m_{AB}$, and for
our length unit, the semimajor axis of the outer companion, $a_{AB}$.  For our standard parameters, the  inner binary has a total mass $m_A$ and is composed of an equal mass binary with $m_{Aa}=m_{Ab}=0.25\, m_{AB} $, semi-major axis $a_A=a_{AB}/20$ and eccentricity $e_A=0$ and an inclination of $i_A=0$ {(coplanar)} relative to outer companions orbit.  The outer companion to the binary has mass $m_B=m_A=0.5\,m_{AB}$ and is in an orbit with an eccentricity of $e_{AB}=0.5$.

More generally, we define the relative mass of inner binary as $f_A=\frac{m_{Ab}}{m_{Aa}+m_{Ab}}$, where $m_{Ab}$ is the smaller of the two masses and so this parameter ranges from 0 to 0.5. The relative mass of companion is $f_B=\frac{m_B}{m_{AB}}$. Since the companion may be smaller or larger in mass than the inner binary this parameter ranges from~0 to~1. 
Our standard 
parameters have $f_A=0.5$ and $f_B=0.5$.
These parameters are in the
stable region for circumtriple systems. The 
system becomes unstable with larger $a_{A}$ and 
$e_{AB}$.  Adopting the Multilayer Perceptron  (MLP)
model from \cite{vynatheya2022} and varying $e_{AB}$ we find it 
is stable for  $e_{AB} \lesssim 0.8$ and
$a_A/a_{AB} \lesssim 0.13$.
%mlp mpdel in /Users/lepp/Desktop/triple-stability-main
We check our ranges of parameters with the MLP model  \citep{vynatheya2022} to avoid unstable regions.  However, the 
transition between stable and unstable is gradual rather than 
abrupt \citep{hayashi2022} and so we have chosen our standard
parameters to be well clear of unstable regions.

\subsection{Test particle orbits around our standard triple star}

We run test particle orbits at radius $r$ around the triple star. 
The test particles can have unstable orbits if they are too close to the AB binary 
\citep{Holman1999,Quarles2020,chen2020}.  We only consider orbits at radii large enough to be stable. We analyse the test particle orbits in the
frame of the AB binary
made up of the companion star orbiting the inner binary.
We characterise the test particle orbit by its inclination and nodal phase angle
relative to this binary.  The inclination of the orbit is 
given by
\begin{equation}
i = \cos^{-1}(\hat{\bm{l}}_{\rm AB}\cdot \hat{\bm{l}}_{\rm p})\,,
\end{equation}
where $\hat{\bm{ l}}_{\rm AB}$ is a unit vector 
in the direction of the AB binary angular momentum and 
$\hat{\bm{ l}}_{\rm p}$ is a unit vector in the direction of the particles 
angular momentum. 
The nodal  phase angle is the angle measured relative to 
the eccentricity vector of the outer binary and is 
given by 
\begin{equation}
        \phi = \tan^{-1}\left(\frac{\hat{\bm{l}}_{\rm p}\cdot (\hat{\bm{l}}_{AB}\times 
    \hat{\bm{e}}_{\rm AB})}{\hat{\bm{l}}_{\rm p}\cdot \hat{\bm{e}}_{\rm AB}}\right) + 90^\circ ,
\end{equation}
\citep{Chen2019,Chen2020e} where $\phi$ is the phase angle and $\hat{\bm{e}}_{\rm AB}$ is the eccentricity vector of the outer binary. 

We run test particle orbits around our standard triple star  that begin
in circular orbits at 
radii of $r=3.5$, 4.5, 5.5  and 6$\,a_{AB}$.  We start with initial 
inclinations in $10^\circ$ increments from $10^\circ$ to $170^\circ$
and with an initial longitude of the ascending node of $90^\circ$.
The resulting orbits are plotted in the $(i\cos\phi,i\sin\phi)$ phase
plane in the left panel of Fig.~\ref{fig:phase}. The right panel
shows the same information but displays the paths of the particles orbital angular momentum 
vector on the unit sphere.

For low initial inclinations, there is a circulating region shown in green, in which the particle angular momentum vector precesses around the binary angular momentum vector. The retrograde circulation region is shown in blue where
the particles angular momentum  vector is orbiting about the negative of the 
binaries angular momentum vector.   
There is a librating region, shown in red, where the particle angular momentum vector precesses around a stationary inclination. 
This stationary inclination for close in particles is at $i=90^\circ$ and aligned with the binary eccentricity vector. 
As the particle moves to larger orbital radii,  the
stationary inclination moves to higher inclinations.  Once the stationary inclination is $>180^\circ$ there 
are no more librating orbits and the particle has similar dynamics to one around a circular orbit binary since it  nodally precesses about the binary angular momentum vector for all inclinations.
This is very similar to the behavior seen in \cite{Lepp2022} 
where we considered test particle orbits about a binary 
which was precessing due to the effects of general relativity.  
Here the behavior of the triple star system is causing a 
similar precession but at a timescale over an order of magnitude
higher.

All the simulations in this paper were run with zero mass 
test particles but to see the effects of a massive particle we
ran select simulations with various mass particles.  The simulations
are essentially unchanged by introducing a particle up to $m_{AB}/1000$ 
(about a Jupiter mass if $m_{AB}\approx 1\, \rm M_\odot$). A
Jupiter mass particle follows the test particle evolution.  Masses
significantly above this mass can change the evolution.  In 
particular large masses in polar orbits induce a precession in
the outer binary in the opposite direction of that caused by
the inner binary and cause the total precession of the 
outer binary to be slower.

\subsection{Critical radius for librating orbits}

%fig2g.pdf comes from figg.pdf in /Users/lepp/Desktop/3star/circumtriple/polarinout/nine
\begin{figure*}
\includegraphics[width=1.99\columnwidth]{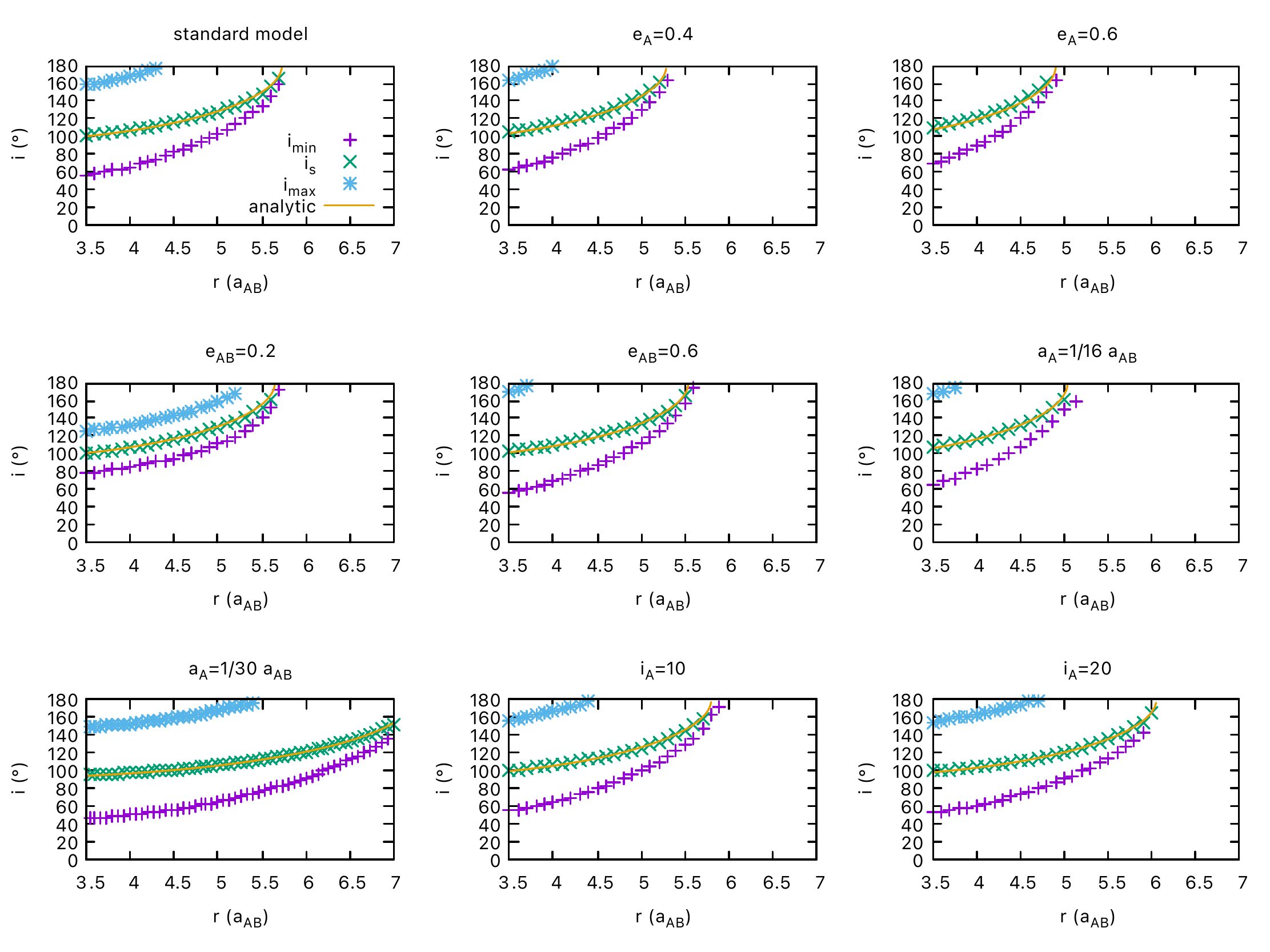}
\caption{The minimum initial inclination ($i_{\rm min}$, magenta) and maximum initial inclination  ($i_{\rm max}$, blue) for librating orbits and the stationary polar inclination (green) for varying $e_A$, $e_{AB}$, $a_{AB}$ or $i_A$ (in degrees) from our standard model.  The analytic
curve for the stationary state is from equation~(\ref{eq:is2}) with $\alpha=2$. The key in upper left panel applies to all nine panels.}
\label{fig:fig2}
\end{figure*}

In Fig.~\ref{fig:fig2} we show the smallest initial inclination for a librating 
orbit, $i_{\rm min}$, the largest initial inclination for a librating orbit, $i_{\rm max}$, and the stationary inclination, $i_{\rm s}$, where the orbit stays at
a fixed inclination with no nodal precession. 
The upper left panel in Figure \ref{fig:fig2} represents our standard
triple star parameters.   There are circulating orbits at low inclinations $i<i_{\rm min}$,  and retrograde circulating orbits  for $i>i_{\rm max}$.  If $i_{\rm max}> 180^\circ$ then there is no retrograde circulating 
region and since the librating orbits occur around $i_{\rm s}$, there are no librating orbits when $i_{\rm s}>180^\circ$. 
The precession of the triple star system causes the stationary 
inclination to move to higher inclinations with increasing 
test particle radius, until it becomes 
greater than $180^\circ$ and then there are no librating orbits.  We 
call this radius the critical radius, $r_{\rm c}$, and it represents the 
maximum radius at which test particles can orbit the outer 
binary in a polar orbit.  For our standard triple star parameters, $r_{\rm c}=5.7\, a_{AB}$.

We now consider the effect of varying the triple star orbital parameters on the test particle orbits. The other panels in Figure \ref{fig:fig2}  take our standard
model and vary one of the parameters.  In the next two 
panels across the top, we vary the eccentricity of the inner 
binary from $e_A=0$ to $e_A=0.4$ and $e_A=0.6$.  The
change in $e_A$ increases the apsidal precession rate of the 
AB binary by about 30\% for $e_A=0.4$ and about 70\% 
for 0.6.  The radius $r_{\rm c}$ then occurs at smaller orbital radii of 5.28 and 4.87$\,a_{AB}$, for $e_A=0.4$ and $e_A=0.6$, respectively.

Next, we vary from our standard case the eccentricity of the 
companion from its value of $e_{AB}=0.5$ to $e_{AB}=0.2$ and 
$e_{AB}=0.6$.  In both cases, $r_{\rm c}$ is
reduced, though the effect  is much weaker than  that seen 
for varying $e_A$.  This is because $e_{AB}$ affects both 
the precession rate of the ascending node of the test particle
and the apsidal precession rate of the AB binary (see Section 
\ref{analytic}).

We then vary the ratio of the semi-major axis of the inner 
binary to the companion from its standard value of $a_A/a_{AB}=1/20$.
For $a_A/a_{AB}=1/16$ we find $r_{\rm c} \approx  5\,a_{AB}$ and for $a_A/a_{AB}=1/30$,
  $r_{\rm c}=7.25 \,a_{AB}$ which is off the range of the plot.  This again 
reflects the change in apsidal precession rate of the binary 
with changing geometry.
Finally, we consider the effect of the inclination of the inner binary relative to the triple star companion. 
We have restricted our simulations to small angle inclinations to avoid 
ZKL oscillations that would introduce additional time variations.  We change our standard model
to have the inner binary's orbit inclined to the orbital plane
of the companion and this increases $r_{\rm c}$.  The critical
radius gets larger as the apsidal precession rate gets smaller. However, we note that the inclination has a weak effect on the critical radius.

%from /Users/lepp/Desktop/3star/circumtriple/polarinout/maxr
\begin{figure*}
\includegraphics[width=2\columnwidth]{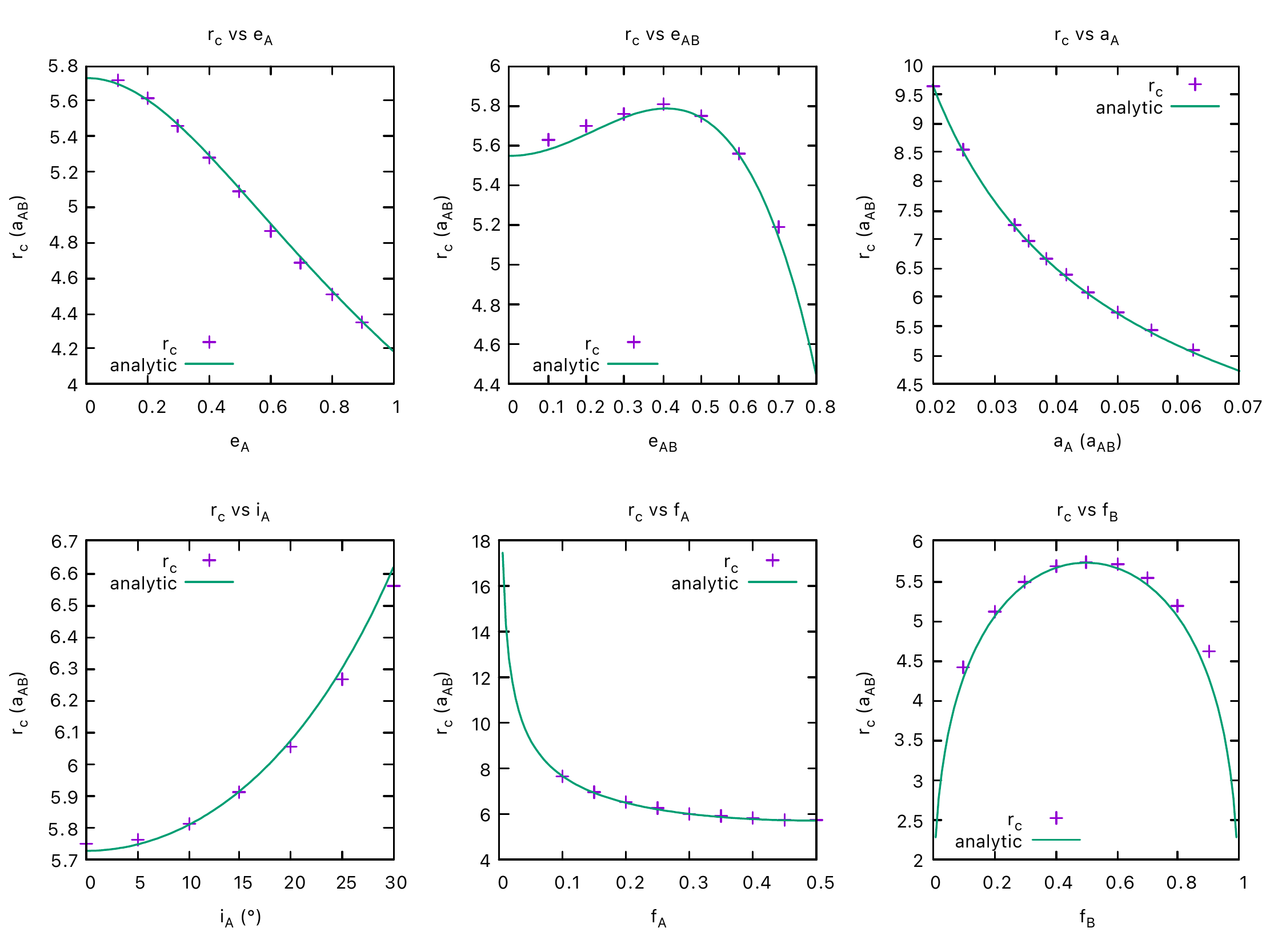}
\caption{The critical radius, $r_{\rm c}$, inside of which there are polar orbits for
varying $e_A$, $e_{AB}$, $a_A$, $i_A$, $f_A$ and $f_B$  from our standard model parameters. The purple crosses show numerical simulations and the green lines show the analytical estimate. The analytic lines only depends on $\alpha$ in the top left plot (since $e_A=0$ everywhere else) and there we take $\alpha=2$.}
\label{fig:fig3}
\end{figure*}

Figure \ref{fig:fig3} shows the critical radius $r_{\rm c}$ as a function of some of the triple star parameters. The crosses show the numerical determination of the radius.  We vary $e_A$, $e_{AB}$, $a_A$,  $i_A$, $f_A$ and $f_B$. 
The critical radius depends on the rate of apsidal precession
of the AB binary as well as on the nodal precession rate of 
the test particle orbit (see the next section).  The faster the apsidal precession the smaller the critical radius. For typical triple star parameters  \citep{Tokovinin2008,Tokovinin2021}, the critical radius is in the approximate range $3-10\,a_{AB}$, unless one of the stars has a much smaller mass than the others.  
The innermost stable orbit for a polar circumbinary test particle is typically around $2-2.5\,a_{AB}$ \citep{chen2020} and so the radial range of stable polar circumtriple orbits may be quite small.
{However, as discussed in Section \ref{sec:params}, some observed triples found with short period inner binaries have more extreme outer to inner semi-major axis ratios that allow the critical radius to extend to more than $80\,a_{AB}$. }

The libration time scale increases with the semi-major axis of the test particle. 
For our standard model, the critical radius outside of which there are no polar orbits is $r_{\rm c}=5.73 \,a_{AB}$.  In this case,  a test particle at $r=5.5 \,a_{AB}$ orbits near the stationary inclination librate about it with a period of about $4000 P_{AB}$, where $P_{AB}$ is the orbital period of the outer binary.  At an orbital radius of $r=4.5 \,a_{AB}$, orbits near the stationary inclination librate with a period of about $800\, P_{AB}$.

\section{Analytical estimation}
\label{analytic}

The stationary inclination occurs where the apsidal precession rate of the binary is equal to the nodal precession rate of the test particle.  We follow \cite{Zanardi2018} to analytically find the stationary inclination based on the quadrupole order expansion of the Hamiltonian. They derived it for the case where general relativity drives the apsidal precession. The precession of  the ascending node of the test particle is given by equation~(4) in 
\cite{Zanardi2018}. For a circular ($e=0$) polar stationary orbit ($\Omega=90^\circ$), the nodal precession rate is

\begin{equation}
\dot \Omega_{\rm s}
= - \frac{m_A m_B k}{m_{AB}^{3/2}r^{3/2}}
\left(\frac{a_{AB}}{r}\right)^2
\frac{3\cos i (1+4e_{AB}^2)}{4},
\end{equation}
where  $k^2$ is the gravitational constant.
We equate this to the rate of change of the 
longitude of the periapsis for the binary, $\dot{\varpi}_{AB}=\dot \omega_{AB}+\dot\Omega_{AB}$, to find the stationary
inclination for the test particle 
\begin{equation}
    i_s =\cos^{-1}\left(
    -\dot{\varpi}_{AB} 
       \frac{4}{3k} \frac{(m_{AB})^{3/2}}{m_Am_B} \frac{r^{7/2}}{a_{AB}^2}
  \frac{1}{(1+4e_{AB}^2)}\right).
    \label{eq:is}
\end{equation}
This formula is general and the apsidal precession rate for the binary could come from general relativity
 \citep[e.g.][]{Zanardi2018}, tidal interactions  \citep[e.g.][]{Sterne1939}
or interactions with a companion star \citep[e.g.][]{Morais2012}.

\begin{figure*}
\includegraphics[width=\columnwidth]{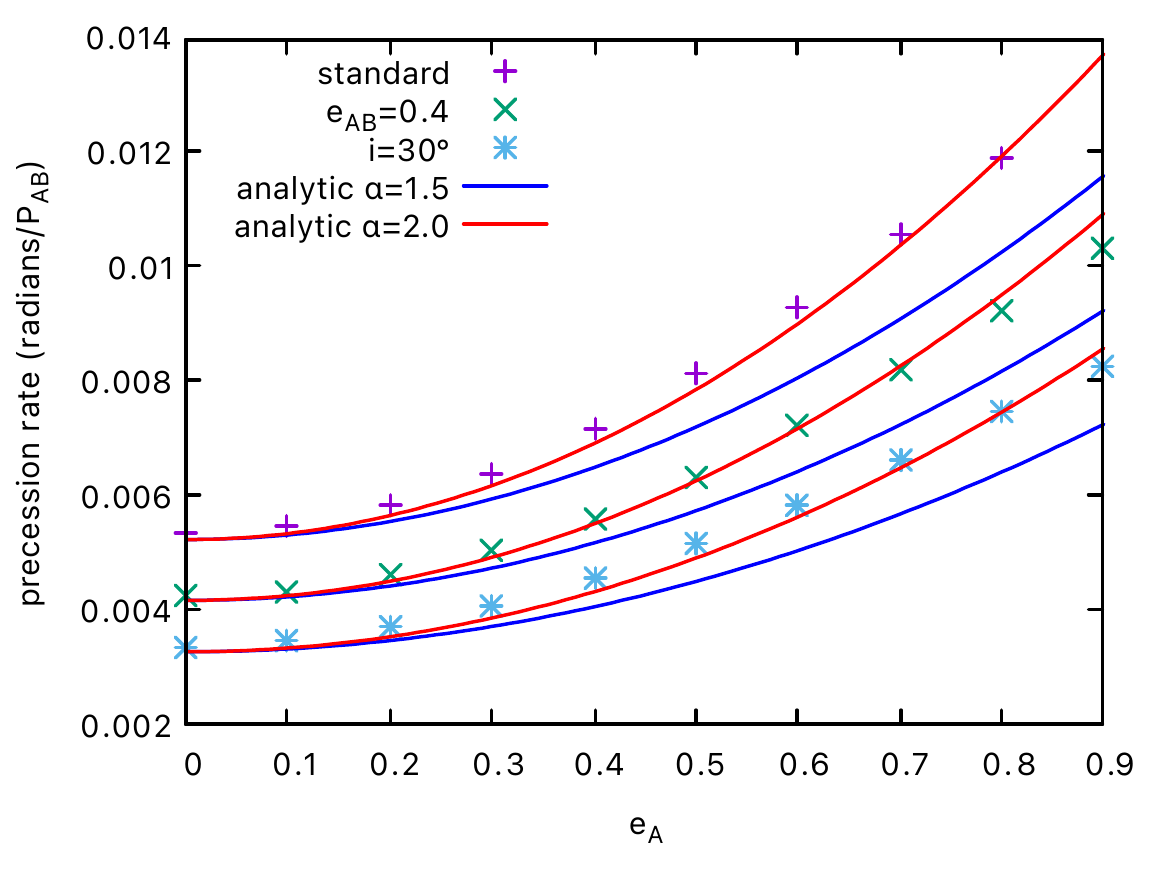}
\includegraphics[width=\columnwidth]{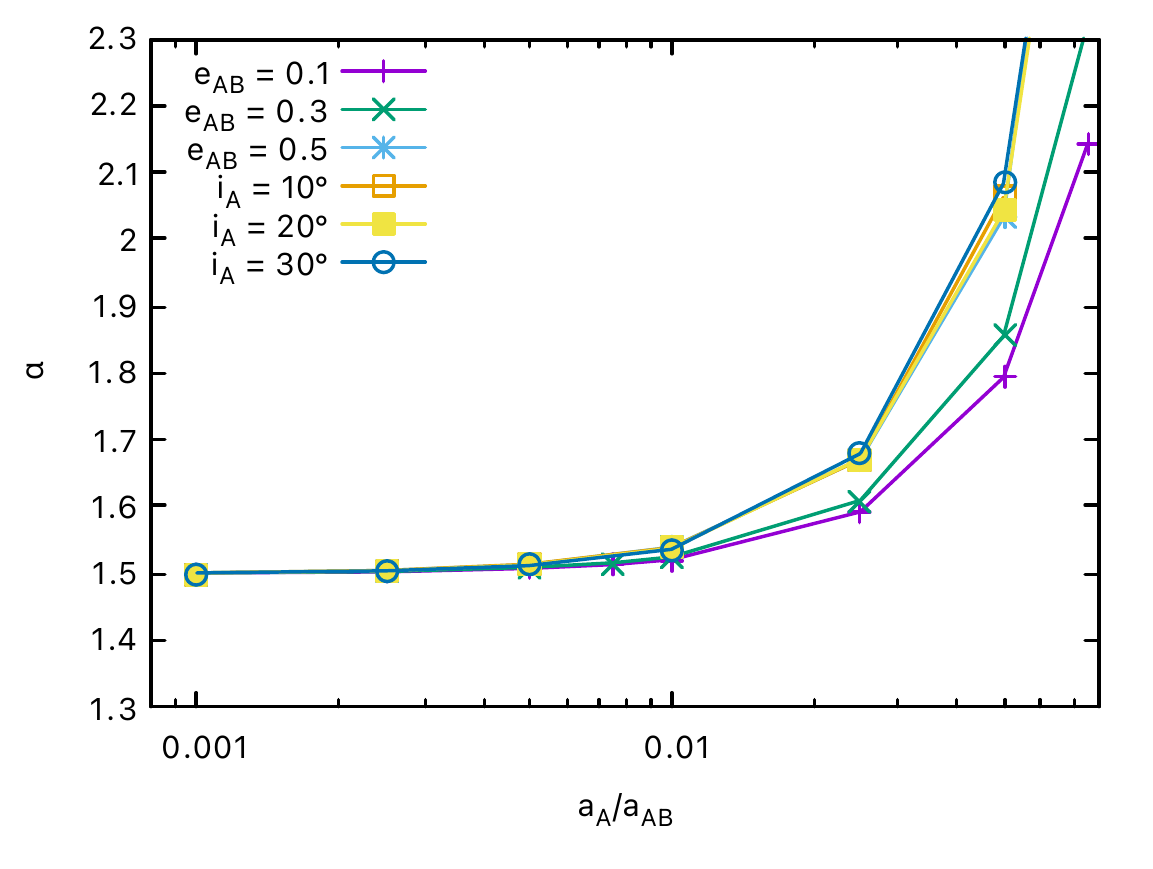}
\caption{ Left: Apsidal precession rate ($\dot{\varpi}$) variation with $e_A$ around our
standard triple star, one with $e_{AB}=0.4$ and one with an inclination
$i_A=30^\circ$.  The analytical fits are shown with  coefficients of both $\alpha=1.5$ and $\alpha=2.0$. Right: The coefficient $\alpha$ in 
front of $e_A^2$ in the precession formula averaged over the values 
of $e_A$ from 0.1 to 0.9.}
\label{fig:alpha}
\end{figure*}

The precession rate of the longitude of the periapsis of 
the companion in a triple in the quadrupole approximation is given by 
\begin{multline}
\dot \varpi_{AB} =
\left(\frac{3 k }{4} \right)
\left(\frac{m_{Aa} m_{Ab}(m_{AB})^{1/2}}{(m_A)^2}\right)\times\\
\left(\frac{a_A^2}{a_{AB}^{7/2}}\right)
\left(\frac{1}{(1-e_{AB}^2)^{2}}\right)F(e_A,i_A),
\label{eq:efit}
\end{multline}
where 
\begin{multline}
    F(e_A,i_A)= (1+\alpha e_A^2)\left(\frac{3\cos(i_A)^2-1}{2}
    \right)\\
    +\frac{15}{4}e_A^2(1-\cos(i_A)^2)\cos(2\omega_A)
    \label{eq:rateei}
\end{multline}
\citep[see equations 25 and 26 in][]{Morais2012},
where $i_A$ is the inclination of the inner binary relative
to the outer binary,
and $\omega_A$ is
the argument of the periapsis of the inner binary measured relative
to the outer binary.   The same rate may also be found for the
co-planar case by adding all the quadrupole terms for $\dot\omega$ 
and $\dot\Omega$ for the outer binary (equations 74 and 
76 in \cite{Naoz2016}), 
the inclination dependence is slightly different as \cite{Morais2012}
approximate the outer binary as the fixed plane. In all our configurations the outer binary carries most of the angular momentum and so this is a good approximation, as seen in Figs.~\ref{fig:fig2} to~\ref{fig:alpha}.  
The first term in equation~(\ref{eq:rateei}) sets the average rate of
apsidal precession and the second term  causes an oscillation
about this average precession rate. 
If $\omega_A$  is an odd multiple of $45^\circ$ then the second term is zero.  In practice, one can ignore the second term if one
wants the average precession over long times or in a time which is centered around an odd multiple of $45^\circ$.
The expression is valid 
so long as $a_A \ll a_{AB}$ and $m_B$ is not much less than $m_A$ meaning that the AB binary has most of the angular momentum and thus the 
AB binary plane is very nearly a fixed plane in the system.

For an inclination
of zero, the function in equation~(\ref{eq:rateei}) simplifies to 
\begin{equation}
F(e_{AB},0)=(1+\alpha e_A^2).
\label{eq:rateei0}
\end{equation}
Here,  the $\alpha=\frac{3}{2}$ factor
given in \cite{Morais2012} works well in
the limit of $a_{AB}\gg a_A$ and we find it works well for ratios of 
$a_A/a_{AB} \lesssim 0.005$.  However,  our standard triple star has $a_A/a_{AB}=0.05$ and so  higher order terms have changed this 
parameter. In  \cite{Naoz2016} it is clear that the octupole terms 
vanish for our standard case of an equal mass inner binary and so
it must be due to even higher order terms \citep[e.g.][]{Yokoyama2003,Vinson2018,de.Elia2019}. We now consider numerical fits for this parameter.

The left panel of Figure \ref{fig:alpha} shows the numerically determined particle precession rates around the triple star as a function of $e_A$. We consider the standard model, the standard model with  $e_{AB}=0.4$ and 
the standard model  with $i_A=30^\circ$.
 To get an accurate precession 
rate, we average the precession rate over  40 periods of the AB binary and over a time with the relative angle of the precession of $45^\circ$, this assures that the second 
term in equation~(\ref{eq:rateei}) averages to zero.

In the right hand panel, we show a numerical determination of $\alpha$ as a function of 
$a_A/a_{AB}$. We take  $e_{AB}= 0.1$, 0.3 and 0.5 with a coplanar binary and  inclinations of 10, 20 and $30^\circ$ with $e_{AB}=0.5$. For each point, we vary
$e_A$ between 0.1 to 0.9 in steps of 0.1 and find for each an exact $\alpha$
that would give that rate relative to the $e_A=0$ rate. We then average all of these.   
 At  
small ratios of $a_A/a_{AB}$ the best fit for $\alpha$ is $3/2$ but close to our standard model $a_A/a_{AB}=0.05$, a much better fit is 
$\alpha=2$.  

We now take $\dot\varpi_{AB}$ from equation~(\ref{eq:efit}) 
and find the stationary inclination with equation~(\ref{eq:is}) to be
\begin{multline}
    i_s=\cos^{-1}\left(
   -  \frac{m_{Aa}m_{Ab}m_{AB}^2}{m_A^3m_B}
     \left(\frac{r}{a_{AB}}\right)^{7/2}
    \times \right. \\
    \left.
    \left(\frac{a_A}{a_{AB}}\right)^2
   \frac{1}{(1+4e_{AB}^2)(1-e_{AB}^2)^2}\, F_e(e_A,i_A)
    \right).
    \label{eq:is2}
\end{multline}
The orange lines in Fig.~\ref{fig:fig2} show the analytic stationary inclination with $\alpha=2$. There is good agreement between this and the numerical solutions.

We find the critical particle orbital radius outside of which there are no polar orbits by setting $i_s=180^\circ$ and solving for $r$ to find
\begin{multline}
    \frac{r_c}{a_{AB}}=\left(
    \frac{m_A^3m_B}{m_{Aa}m_{Ab}m_{AB}^2}
     \left(\frac{a_{AB}}{a_A}\right)^{2}   \times \right. \\ \left.
    \frac{(1-e_{AB}^2)^2(1+4e_{AB}^2)}{F_e(e_A,i)}
    \right)^{2/7}.
    \label{eq:rc}
\end{multline}
Our standard parameters have  $r_c/a_{AB}=5.73$, in agreement with the top left panel of Fig.~\ref{fig:fig2}. More generally we find
\begin{equation}
    \frac{r_c}{a_{AB}}=5.73\, \frac{M \,A\, E}{F^{2/7}},
\end{equation}
where $M$, $A$ and $E$ are scaling functions for 
the radius in terms of mass, semi-major axis, and eccentricity of
the companion which have been normalized to one for our standard parameters.  The scaling with masses is
\begin{align}
    M(m_{Aa},m_{Ab},m_B) & = \left( \frac{m_A^3m_B}{m_{Aa}m_{Ab}m_{AB}^2} \right)^{2/7}\\
    & = \left(\frac{(1-f_B)f_B}{(1-f_A)f_A}\right)^{2/7},
    \label{eq:mass}
\end{align}
 with semi-major axis is
\begin{equation}
    A(a_A,a_{AB}) =  \left(\frac{a_{AB}}{20\, a_A}\right)^{4/7},
\end{equation}
with the companion eccentricity is
\begin{equation}
        E(e_{AB}) =
        \left( \frac{8 (1-e_{AB}^2)^2(1+4e_{AB}^2)}{9} \right)^{2/7}.
\end{equation}
The green lines in Fig.~\ref{fig:fig3} show this analytic solution for the critical radius. We see that there is good agreement between the numerical and analytic solutions. 

\section{Discussion and Conclusions}
\label{conc}

Misaligned circumbinary test particle orbits around an eccentric binary undergo nodal precession either about the binary angular momentum vector ($i=0^\circ$) or about the stationary polar inclination that is aligned to the binary eccentricity vector ($i=90^\circ$).  The orbit type depends on the initial particle inclination and the binary eccentricity but it does not depend upon the particle semi-major axis. With $n$-body simulations and analytic methods we have investigated the dynamics of circumtriple particle orbits. For close in particles, the polar inclination is $90^\circ$ and the orbits around the triple star are similar to those around the outer binary with the inner binary replaced by a single star. However, with a hierarchical triple star, the inner and outer binaries  undergo apsidal precession and this leads to an increasing  polar stationary inclination with increasing particle semi-major axis. {There is a critical radius $r_{\rm c}$ outside of which there are no polar orbits, only circulating orbits that precess about the binary angular momentum vector. We find for typical parameters that the critical radius is in the approximate range $3-10$ times the outer binary semi-major axis. }
 Therefore, polar circumtriple orbits {typically} exist only relatively close to a triple star.  But for some observed  shorter period inner binaries ($< 1y$),  the ratio of the outer to inner semi-major axis is quite large \citep{Tokovinin2021}. {In such cases, the  circumtriple orbits can  occur at relatively large distances from the outer binary.}

A low-mass circumtriple disk can undergo similar behaviour to the particles, but the radii of the disc communicate with each other allowing solid body precession. Therefore, a disk with an outer radius {larger} than $r_{\rm c}$ could reach a polar state. However, because $r_{\rm c}$ can be only a few times the outer binary separation, even if a disk began with an outer radius smaller than $r_{\rm c}$, it may quickly spread out beyond this, depending upon the disk viscosity. This suggests that a polar circumtriple disk could form, although it may be the inner part of a broken disk. If $r_c$ is small, communication through the disk may instead lead the outer parts to dominate the behaviour and the disc to move towards coplanar alignment. These effects should be investigated in future work. 

There are two triple star systems that may have planets orbiting them, GG Tauri A \citep{Phuong2020a} and 
GW Ori \citep{Bi2020,Smallwood2021}.  The GG Tauri~A system consists of
three stars \citep{DiFolco2014} with $m_B=0.6\,\rm M_\odot$, $m_{Aa}=0.38\,\rm M_\odot$ and  $m_{Ab}=0.3\,\rm M_\odot$.  The outer binary
semimajor axis  is estimated to be
$a_{AB}=36 \,\rm au$ and  inner binary semi-major axis is about $a_A=5.1 \,\rm au$.
The other 
orbital parameters are uncertain, but we can estimate $M\approx 1$ and  $A\approx 0.55$ and $r_{\rm c}=3.2\, a_{AB}= 113\,\rm  au$.  The disk around the triple extends from $r=180\,{\rm au}\approx 5\,a_{AB}$ to $800 \,\rm au$ and the proposed
planet is at about $230\,\rm au$. 
The 
second system, GW Ori, has the triple star parameters 
$m_{Aa}=2.47\,\rm M_\odot$, $m_{Ab}=1.43\,\rm M_\odot$ and  $m_{AB}=1.36\,\rm M_\odot$, 
$a_A =1.2\,\rm au$, $a_{AB}=8.89\,$au,  $e_A=0.069$ and
$e_{AB}=0.379$ \citep{Kraus2020}. These
give $A=0.56$, $M=0.95$, $F=1$, $E=1.05$ and we find $r_{\rm c}= 3.0 \,a_{AB}= 28.4 \,\rm au$.  The observed disk is at $r>36\,{\rm au}\approx 4\, a_{AB}$ and the proposed planet is at $r=100\,\rm  au$.  Again the disk and the planet are well outside the critical radius.

 For both of these observed circumtriple systems, the inner edge of the disk  and the orbits of the potential planets are larger than $r_{\rm c}$. This suggests that the dynamics of the planet and disk are similar to that around a circular orbit binary. A coplanar circumbinary disc is truncated at $2-3$ times the binary separation \citep{Artymowicz1994} and the cavity size decreases with increasing tilt of the disc \citep{Miranda2015,Lubow2018,Franchini2019}. Therefore the inner truncation radius of both of these disks are larger than would be predicted for a circumbinary disk.  This could be a result of the triple star effects described in this work that limit the disc to be in $r>r_{\rm c}$. The tidal truncation of a misaligned circumtriple disk should be investigated in future work. 

\ 

%\begin{acknowledgements}
We thank an anonymous referee for useful comments that improved the manuscript. We acknowledge support from NASA through grants 80NSSC21K0395 and 80NSSC19K0443.  SHL thanks the Institute for Advanced Study for visitor support.
%\end{acknowledgements}

\bibliographystyle{aasjournal}
\bibliography{ct} 

\end{document}